# Model-based Product Quality Evaluation with Multi-Criteria Decision Analysis

*Adam Trendowicz, Michael Kläs, Constanza Lampasona, Jürgen Münch*

Fraunhofer Institute for Experimental Software Engineering
Fraunhofer-Platz 1, 67663 Kaiserslautern, Germany

{adam.trendowicz, michael.klaes, constanza.lampasona, juegen.muench}
@iese.fraunhofer.de

*Christian Körner, Matthias Saft*

Siemens AG, Corporate Technology

Otto-Hahn-Ring 6, 81739 Munich, Germany

{christian.koerner, matthias.saft}@siemens.com

*Abstract:*

*The ability to develop or evolve software or software-based systems/services with defined and guaranteed quality in a predictable way is becoming increasingly important. Essential - though not exclusive - prerequisites for this are the ability to model the relevant quality properties appropriately and the capability to perform reliable quality evaluations. Existing approaches for integrated quality modeling and evaluation are typically either narrowly focused or too generic and have proprietary ways for modeling and evaluating quality. This article sketches an approach for modeling and evaluating quality properties in a uniform way, without losing the ability to build sufficiently detailed customized models for specific quality properties. The focus of this article is on the description of a multi-criteria aggregation mechanism that can be used for the evaluation. In addition, the underlying quality meta-model, an example application scenario, related work, initial application results, and an outlook on future research are presented.*

## 1 Introduction

Developing software or software-based systems/services with plannable quality requires appropriate means for understanding, defining, controlling, predicting, and improving quality properties. For this purpose, reliable models of software quality and associated quality evaluation methods are required. Objective measurement and evaluation of software quality can be seen as a fundamental basis for these instruments.

Quality models must be easily adaptable to a particular organization, allow for quantitative statements regarding quality, and support decision-making from various per-spectives and for different purposes. For example, they should support defining and communicating quality requirements, monitoring and controlling quality, and predicting quality. Moreover, quality models should be associated with a systematic method for evaluating quality. Such methods should replace frequently used ad-hoc quality evaluation approaches based on human judgment. In order to gain acceptance, they must be intuitive and, at best, reflect human reasoning process.

Existing approaches for quality modeling are typically either narrowly focused or too generic and have proprietary ways for modeling and evaluating quality. Shortcomings can be seen with respect to systematic operationalization, adaptation support, and limitation to a specific set of application purposes.

In this article, we propose a flexible quality modeling and evaluation approach. The approach is based upon a well-defined quality meta-model developed by the Quamoco consortium [19] and uses multi-criteria decision analysis (MCDA) techniques. Moreover, we embedded the method within a continuous improvement framework. In the following sections, we first provide an overview of related work in the area of quality-model-based product evaluation and the application of MCDA methods in the context of product quality evaluation. Next, we present the quality model structure on which the proposed evaluation is based. Then, we present our own idea: how MCDA methods can be integrated with the model structure to provide the possibility of repeatable quality evaluations. An initial application of our idea is described and results are discussed. Finally, we summarize our current work and sketch planned research directions.

## 2 Related Work

In the literature, several schemata for classifying quality models have been proposed. One example is to classify quality models by using so-called landscapes [11]. Here, one of the classification dimensions distinguishes nine major application purposes for quality models (i.e., specify, measure, monitor, assess, control, improve, manage, estimate, and predict). Quality models classified as supporting the purposes "assess" or "control" should provide a means for quality evaluation. In this context, "assess" means *quantifying and measuring a concept in order to compare it to defined evaluation criteria for the purpose of checking the fulfillment of these criteria*.

The authors of this article have classified many different quality models consisting of different conceptual constructs and supporting different purposes [7]. Figure 1 presents a summary of the classified product quality models according to the dimensions *purpose* and *quality focus*. 16 quality models were found whose purpose is to "assess" and two that have "control" as a major application purpose. Of these, 13 focus on general quality, one on defects, three on maintainability, and one on safety. These models provide many different means for quality evaluation. Jin et al. [10], for example, use a fuzzy approach for evaluating qualitative as well as quantitative indicators.

|          | General | Defects | Dependability | Functionality | Maintainability | Portability | Reliability | Safety | Usability |
|----------|---------|---------|---------------|---------------|-----------------|-------------|-------------|--------|-----------|
| Specify  | 20      | 0       | 4             | 0             | 1               | 2           | 0           | 4      | 1         |
| Measure  | 16      | 1       | 0             | 2             | 2               | 0           | 0           | 0      | 1         |
| Monitor  | 1       | 2       | 0             | 0             | 0               | 0           | 1           | 0      | 0         |
| Assess   | 13      | 0       | 0             | 0             | 2               | 0           | 0           | 1      | 0         |
| Control  | 0       | 1       | 0             | 0             | 1               | 0           | 0           | 0      | 0         |
| Improve  | 7       | 0       | 2             | 0             | 2               | 0           | 1           | 0      | 0         |
| Manage   | 4       | 1       | 0             | 0             | 0               | 0           | 0           | 0      | 0         |
| Estimate | 2       | 7       | 0             | 0             | 0               | 0           | 0           | 0      | 0         |
| Predict  | 4       | 5       | 0             | 0             | 1               | 0           | 3           | 0      | 0         |

**Figure 1**: Quality model landscape excerpt

Ciolkowski and Soto [5] use interpretation rules to map metrics to interpretations. They apply a four-value scale, which uses an association with traffic lights (red=bad, yellow=satisfactory, green=excellent, and black=unacceptable). After interpretation, they aggregate several interpretations into a single one using rule-based approaches.

A current survey indicated that aggregation of quality indicator results is a common activity in practice [18]. However, it is not predominant. One cause of this could be that common quality models do not sufficiently support aggregation. One example of the use of quality assessment in practice is presented in [13]. SAP measures Key Performance Indicators (KPIs) for software quality and aggregates the results into a quality index.

MCDA techniques present a promising way for evaluating software quality and have actually already been applied by several authors. Formal decision support techniques have been used to evaluate and choose candidate software architectures [17], COTS components [4, 16, 14], and domain-specific software applications [1, 2]. The techniques applied included both compensatory [2, 4, 16, 17] and non-compensatory [4, 14] techniques, with the Analytic Hierarchy Process (AHP) being the most commonly used compensatory approach. Quality models used for evaluation purposes typically either adopted ISO9126 [9] directly or their structure is conformant to ISO – thus taking over drawbacks of the standard.

## 3   Quality Meta-Model

In order to perform a systematic evaluation of the quality of a software product, we need a quality model that allows us to describe what product quality means for us, what contributes to it, and how we can quantify and evaluate it.

This means that a quality model to be used in the systematic evaluation of product quality requires certain types of elements (e.g., relevant quality aspects, measures, evaluation rules, etc.) [11]. In addition, in order to be able to provide a quality eval-uation method on a level of abstraction that allows its practical application, we have to assume a certain kind of structure of the quality model. Such a structure has to describe, for instance, which information is provided by which element of the model and which kinds of relationships are assumed between the model elements. A good way to describe such information is to use a *meta-model* [8].

Unfortunately, many common quality models do not explicitly define their underlying meta-model. Moreover, several quality models do not provide the model elements required to perform a systematic quality evaluation with repeatable results. For instance, the ISO9126 standard [9] does not provide explicit model elements for capturing thresholds for the quality evaluation or for describing aggregation rules for combining the evaluation results of different elements. Therefore, the evaluation method presented in this paper is based on a more detailed meta-model than the one provided, for example, by ISO9126.

The meta-model we use in this paper was developed in the Quamoco project by a consortium of research and industry partners with the explicit objective of supporting the description and evaluation of product quality [19]. In the following subsections, we briefly describe the general concepts behind this meta-model and illustrate its different parts with a simple example.

## 3.1   General Concepts

Figure 2 gives an overview of the meta-model. In general, we can distinguish two levels of the meta-model, the definition level and the application level. The elements on the *definition level* allow defining quality and describing how to evaluate it. They are instantiated when the quality model is defined and are independent of the specific application for a product. The element on the *application level* allows using the quality model definition elements and applying them to a specific software product by gathering and analyzing information according to the rules provided by the definition level. The definition level can again be separated into two parts, the specification part and the evaluation part. The elements of the *specification part* allow defining the quality of software products qualitatively and the elements of the *evaluation part* define software quality quantitatively and explain how to evaluate the quality.

Finally, we can distinguish three kinds of conceptual parts in the meta-model [11]:

The elements on the right-hand side deal with the *quality focus,* meaning the quality (aspects) of interest in our model, such as product reliability, maintainability, or the product quality in general.

The *variation factor* part on the left-hand side captures information on the factors, more specifically, properties of the product that are assumed to influence the quality of interest (e.g., the complexity of the product or its level of documentation).

The middle part considers the *relationship* between the variation factors and the quality focus [11]. This allows to explicitly describe, for instance, why and in which way a certain factor impacts a certain aspect of the quality focus.

## 3.2 Illustration of the Meta-Model's Elements

We provide in this section a brief overview of the elements of the meta-model based on a very simple example (Figure 2). A more detailed description and a more exhaustive example can be found in [19].

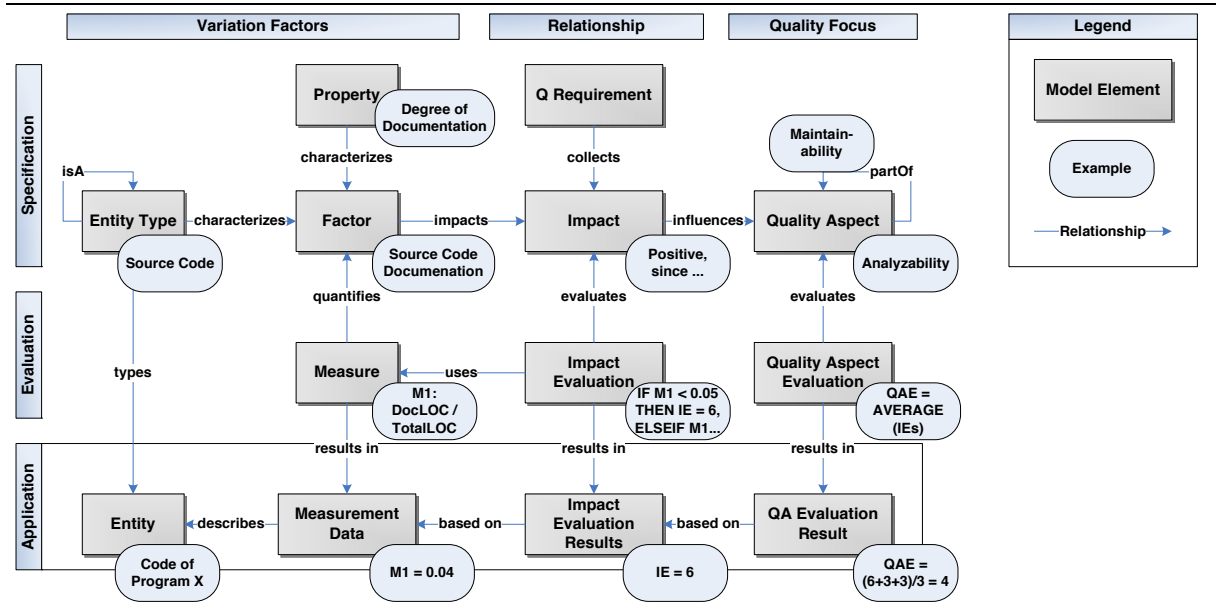

**Figure 2:** Quality meta-model with examples for model elements

First, we describe and illustrate the quality *specification part* of the meta-model. In our example, we are interested in evaluating the maintainability of a product. Therefore, maintainability represents the root node of our quality aspect tree. Maintainability may be refined into sub-aspects such as analyzability, changeability, stability, etc. In the following, we focus on analyzability, but these descriptions also apply to other sub-aspects. There are several factors that influence a product's analyzability. One such factor is, for instance, the degree of source code documentation. As we can see in this example, a factor consists of two components: (1) the type of entity it describes (source code) and (2) the property of this entity that has an influence on a specific quality aspect (degree of documentation).

A sufficient degree of documentation makes the source code easier to understand and, therefore, increases the analyzability of the product. The information regarding the direction of the impact (positive or negative) and its justification is captured by the *impact* element. It establishes the relationship between the factor and the influenced quality aspect in a qualitative way.

Next, we describe and illustrate the *quality evaluation part* of the meta-model by extending the example. In order to determine the analyzability of the source code, we have to define at least one *measure* that allows us to quantify it. Such a measure may be the ratio between the number of comment lines and the overall number of lines of code (M1). We choose this measure, which provides as a result a percentage rate between 0 and 1, for reasons of simplicity. In order to evaluate the factors' impacts on a specific quality aspect, we need to map the range of possible measurement results to the evaluation scale. Such a scale could be, for example, the grades between 1 (best) and 6 (worst). This mapping is provided by the *impact evaluation* element. Finally, multiple factors may influence a quality aspect, so we need to aggregate the evaluation results of different impact evaluations in order to obtain one evaluation result for a specific quality aspect. In our example, we may have two other factors that influence analyzability. Further, we assume that all are considered equally important; therefore, the corresponding *quality aspect evaluation* element simply provides the rule for computing the average of the three impact evaluation results.

Finally, we describe and illustrate the *application level* of the meta-model. Let us assume that we have a product X that we want to evaluate with respect to its analyzability. In order to do this, we first have to identify the *entities* we have to measure (in this simple case, the source code of product X). Then, we have to perform the measurement and collect the *measurement data* (M1 = 0.04). Next, we use the mapping provided by the impact evaluation to get the corresponding grade (in this case, 6). Finally, the impact evaluations of all factors that influence analyzability are aggregated by the rules provided by the corresponding quality aspect evaluation element. In our case, we would simply calculate the average of the impact evaluation results 6, 3, and 3 and arrive at a final rating of 4 for the analyzability of the product.

## 4 Quality Evaluation Approach

### 4.1 General Quality Assessment Process

Like any other process or technology within a particular organization, quality evaluation should also be the subject of continuous improvement. For this purpose, we propose integrating it into the well-known and widely accepted quality improvement paradigm (QIP). QIP defines six fundamental steps (Figure 3) that are implemented at two different levels.

The quality evaluation method includes activities at both of these levels:

- *Organization level*: At this level, the standard quality model and the associated methods and processes – including the quality evaluation method – are adjusted for a certain application context and are improved continuously.
- *Product level*: At this level, the quality model defined and maintained at the organization level is applied for assessing, analyzing, and improving the quality of concrete software products. Note that, in practice, multiple product-level improvement cycles can run in parallel to the organization-level improvement cycle. For a detailed description of the QIP-based quality evaluation process, please refer to [12].

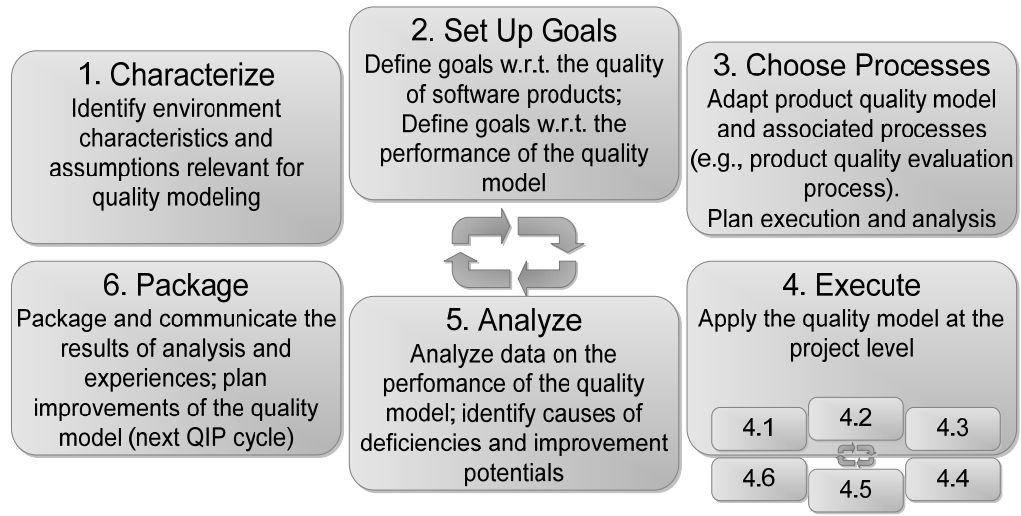

**Figure 3**: General quality evaluation process

## 4.2 MCDA-based Quality Evaluation

This section provides the principles of the quality evaluation method. We based the method on the well-known hierarchy-based MCDA methods where decisions are made via a tree structure of criteria and sub-criteria. In the context of quality evaluation and the Quamoco model, hierarchy of criteria corresponds to hierarchy of factors, impacts, and quality aspects. Similar to an MCDA tree, measurement data are provided in the leaves of the quality model, and decisions – such as the quality evaluation in this case – are made at its root.

The quality evaluation approach was inspired by two particular MCDA methods: AHP [15] and AvalOn [16]. The AvalOn method actually follows the principles of AHP, but leverages the drawbacks of AHP by applying a weight rebalancing algorithm. In contrast to AHP, for instance, it allows for any modification (add, delete) of the set of alternatives while maintaining consistency in the preference (dominance relation) among the alternatives.

**Hierarchy-based MCDA**

The objective of MCDA [3] can be defined as supporting a decision maker in obtaining objective information regarding the preference among a set of alternatives based on multiple decision viewpoints (criteria). A preference among the *alternatives* (decision variants) $a \in A$ is defined upon the decision *criteria* $g \in G$, where the j-th criterion $g_j$ represents the alternative's attribute (characteristic) that is relevant from the perspective of a decision maker for deciding about the preference among the alternatives. On the level of a single criterion $g_i$ the preference of an alternative $a$ - $Pref_i(a)$ - is typically represented by the value $g_i(a)$ (so-called *score*) assigned to an attribute according to a *metric* associated with the criterion and an appropriate measurement procedure (e.g., expert judgment). In many MCDA problems, it is convenient to rep-resent the preference of an alternative *a* indirectly by a so-called *value function V*. A value function $V: A \to \mathbf{R}$ rationalizes a relation of preference (dominance) on a set of alternatives *A*, that is, for every pair of alternatives $x, y \in A$, $V(x) \leq V(y) \Leftrightarrow x \preceq y$. This implies that the preference (dominance) relation is complete and transitive.

In the context of multiple decision criteria $g \in G$, a decision maker may perceive these as having different *importance*, which is typically reflected by associating with each criterion $g_j$ a *numerical weight* $w_j$. In fact, the weight on a particular criterion represents two aspects:

- the range in which alternatives may differ on this particular criterion, and
- how much that difference matters to a decision maker

For example [6], temperature measured on both Fahrenheit and Celsius scales can be normalized to the 0 to 100 scale. However, in the case of Celsius, such a normalized scale covers a greater range of temperature because a Celsius degree represents nine fifths more of a temperature change than a Fahrenheit degree. Equating the units by assigning appropriate weights to each Celsius and Fahrenheit factor is formally equivalent to judging the relative importance of each factor, so with the right weighting procedure, the process is meaningful to those making the judgments.

In the presence of weights, the value-based preference for an alternative *a* on the i-th criterion $g_i$ can, for example, be defined as a product (1) of the alternative's score $g_i(a)$ and the criterion's importance $w_i$.

$$Pref_i(a) = V_i(a) = w_i \cdot g_i(a) \tag{1}$$

For *k* multiple decision criteria, the total preference of an alternative *a* can be defined using an additive value function, which assumes that the total value of an alternative is the sum of particle utilities on all *k* criteria (2).

$$Pref(a) = \sum_{j=1}^{k} Pref_j(a) = \sum_{j=1}^{k} w_j \cdot g_j(a) \tag{2}$$

An additive value function represents so-called compensatory approaches, in which mutual compensating of "negative" and "positive" values on all *k* criteria is allowed. In non-compensatory approaches, such mutual neutralization is not allowed.

In a hierarchy-based decision problem, a criterion $g_i$ can be further decomposed into *k* sub-criteria, which may further be decomposed into their sub-criteria, and so on, eventually creating a tree structure of criteria. In this case, an additive value function (2) may be applied for each node (criterion) in the tree to compute its total value (preference) over all its *k* sub-nodes (sub-criteria). In the context of additive value functions, both *V(a) = Pref(a) ∈ [0, 1] and w(a) ∈ [0, 1]* for all criteria and sub-criteria in the tree.

**Constraints and Assumptions**

Analysis techniques that are potentially useful for quality evaluation purposes require various conditions to be met in order to be applied and to provide valid results. In this section, we present the most relevant constraints (C) considered while developing the quality evaluation approach and corresponding assumptions (A) made to simplify the development of an initial evaluation approach. We are aware that some of the assumptions are typically not met in real situations. Therefore, in the next step of our research, we will focus on evolving a more robust quality evaluation approach that deals with typical constraints for real software development contexts.

*Quality model architecture*

*C1*: The quality assessment method is associated with and depends on the architecture of the quality model used. In our research, the quality assessment method is bound to the Quamoco quality meta-model [19].

*A1*: We assume that the quality evaluation method is applicable and provides valid outcomes for any specific quality model compliant to the Quamoco meta-model.

*Independence of quality model elements*

*C2*: The analysis techniques we use for evaluating quality require the input variables to be independent. For example, statistical methods often require variables to be causally independent, whereas MCDA techniques often require criteria to be preferentially and trade-off independent [3].

*A2*: We assume that elements of a quality model that are considered in a single node of a quality model are already causally and preferentially independent.

*Mutual compensation of quality model elements*

*C3*: An important aspect of quality assessment is mutual compensation of the "negative" and "positive" values of aggregated quality aspects.

*A3*: We allow for mutual compensation of quality aspects. In our case, mutual compensation is supported by a consistent scale on which quality aspects are evaluated, namely, a 6-grade ordinal scale (German school grading).

*Acceptable measurement scales*

*C4*: The applicability of certain data analysis techniques depends on the type of scale that input data are measured on.

*A4*: In the initial version of the quality assessment method, we make the following assumptions:

The "Measure" and "Factor" elements of the quality model are measured using any measurement scale;

The "Impact" and "Quality Aspect" elements of the quality model are quantified on a 6-grade ordinal scale, where 1 refers to the best- and 6 to the worst-case evaluation.

*Quality of input data*

*C5*: The applicability of certain data analysis techniques depends on the quality of the input data in terms of their format, validity, and completeness.

*A5*: We assume that the quality of the measurement has been assured before the quality evaluation method is applied. Regarding these data, we expect:

- *Validity* – Data already have the format required by the techniques used in the quality evaluation approach.
- *Integrity* – Data are consistent and correct.
- *Completeness* – Data are complete, that is, data for all measures defined within the quality model are available on the input of the quality evaluation method.

*Uncertainty of information*

*C6*: In industrial practice, information is often uncertain. Therefore, the quality assessment method should handle both certain and uncertain information. Yet, handling uncertain information typically requires involving complex data analysis techniques.

*A6*: Involving complex analysis techniques for handling information uncertainty will make it difficult to verify the core, MCDA-specific elements of the quality evaluation method. In order to isolate the impact such techniques may have on the validity of the evaluation method, we assume that up-to-date and clear information is provided – thus there is no need for handling uncertainty.

**Basic Activities of Quality Evaluation**

In this paper, we propose a quality evaluation approach in which hierarchy-based MCDA techniques are applied to a Quamoco quality model (Chapter 3) for assessing the quality of a software product. In the quality evaluation method, the decision alternative is represented by the software product under evaluation and the criteria are represented by the elements of the quality model structure: factors, impacts, and quality aspects.

Quality evaluation consists of three basic activities: weighting, aggregation, and evaluation.

*Weighting* assigns numerical weights to elements of the quality model structure. In principle, any systematic technique used in the MCDA domain can be employed for weighting quality model. We propose using the pair-wise comparison technique as defined by Saaty [15]. It does not require any quantitative data and has proved over many years to work well in a number of applications. Weighting is applied locally for each node in the model in that all its $k$ direct sub-nodes are compared pair-wise and weights $w_j$ are obtained such that $w_j \in [0, 1]$ and all $k$ weights sum up to 1. Local weights are assigned throughout the quality model in a top-down manner, beginning from its root down to leaves. In order to leverage the drawbacks of applying pair-wise comparisons to a hierarchy of criteria, we follow local weighting by using a weight rebalancing procedure. Detailed specification of the weight rebalancing algorithm is beyond the scope of this paper and can be found in [16].

*Aggregation* implements the concept of value function. For software product *a,* aggregation computes its preference value *Pref(a)* in a certain node of the quality model based on the values $g_j$ and the weights $w_j$ associated with all its direct sub-nodes using an additive value function (2).

*Evaluation* interprets the results of aggregation to a decision maker by mapping them onto a mode-intuitive evaluation scale. For example, less intuitive outcomes of the aggregation (*Pref(a)* $\in$ *[0, 1]*) can be mapped onto an 6-grade ordinal scale, where 1 and 6 mean best and worst value, respectively. Evaluation is actually optional and can be applied already on the root node of the quality model for evaluating the total quality of a product *a*. Evaluation requires prior definition of an appropriate *evaluation function*.

In the quality evaluation method, not all activities are applicable to all elements of the quality model. Table 1 briefly summarizes which evaluation activities are allowed for particular elements of the Quamoco quality model.

| Model Element | Quality evaluation activity | | |
|---|---|---|---|
| | Weighting | Aggregation | Evaluation |
| Measure | n/a | n/a | n/a |
| Factor | No | Yes[a] | No[b] |
| Impact | Yes | Yes | Yes |
| Quality Aspect | Yes | Yes | Yes |

[a] Base measures can be "aggregated" into derived measures
[b] Factor represents an abstract property or derived measure

**Table 1**: Applicability of basic evaluation activities

**Evaluation Procedure**

The quality evaluation procedure (Figure 4) consists of six basic steps.

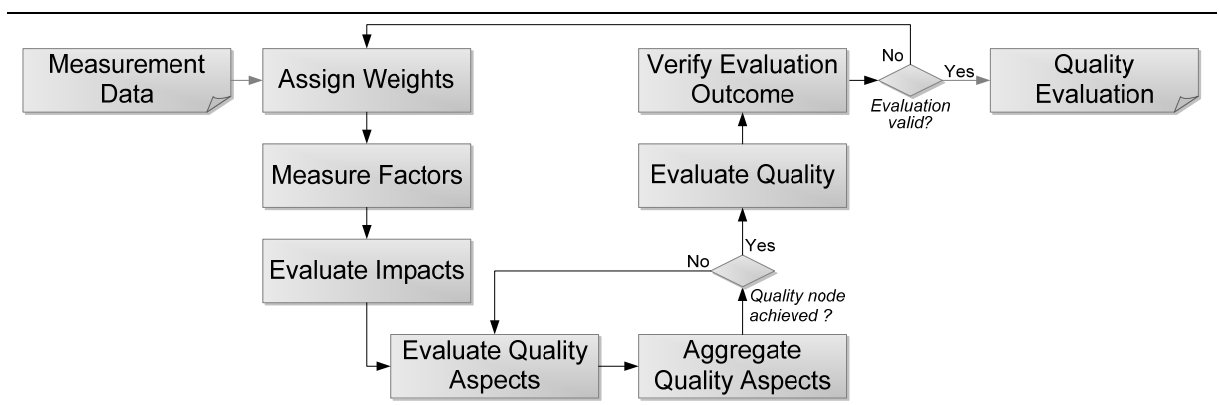

**Figure 4**: Quality evaluation procedure

***Step 1: Assign weights***. Weights are assigned to impacts and quality aspects in a quality model.

***Step 2: Measure factors***. For each factor, the value is determined based upon associated measures:

If the factor represents a derived measure, then its value is computed using the associated base measures; e.g., the code defect density factor is computed by dividing the number of defects found in the code by the number of lines of code.

If the factor represents, i.e., is associated with a single measure, then it assumes the value of this measure.

If the factor represents an abstract property and there are several measures associated with it, then computing its value is optional. For example, the factor *code complexity* can be associated with two measures: McCabe's Cyclomatic Complexity and Halstead's Vocabulary. In this case, we may optionally determine an abstract value of code complexity and use this number later on for evaluating associated quality aspects.

***Step 3: Evaluate impacts***. The impact of each factor on each associated quality aspect is evaluated. The value of the factor or – if it was not computed - the values of associated measures is used for evaluating the factor's impact on the quality aspect.

***Step 4: Evaluate quality aspects***. Quality aspects are evaluated in an iterative process based on the values of associated sub-aspects and factors (via appropriate impacts). Evaluation of each quality aspect consists of two activities:

*Aggregate*: The values of associated sub-aspects and impacts are normalized to the [0, 1] range and are aggregated by means of a value function (2).

*Evaluate*: The outcome of the aggregation can optionally be evaluated using an evaluation function.

***Step 5: Evaluate quality***. Product quality is evaluated based upon the values of associated quality aspects. The evaluation is done analogically to step 4.

***Step 6: Verify evaluation outcome***. The quality evaluation is reviewed. In a simple case, the quality evaluation can be compared to direct expert judgment (common sense). MCDA suggests performing a sensitivity analysis 3, in which the impact of changes to model parameters (e.g., weights, evaluation functions) on the stability of quality evaluation is investigated.

## 4.3   Example Quality Evaluation

In this section, we provide an example that illustrates the core steps (2 to 5) of the quality evaluation procedure we propose in this paper. We exclude steps 1 and 6 because they would require much additional description. We assume that the weights *w* are already provided and balanced. Figure 5 illustrates a simple quality model and provides project measurement data used in our example. Table 2 provides project measurement data used in our example.

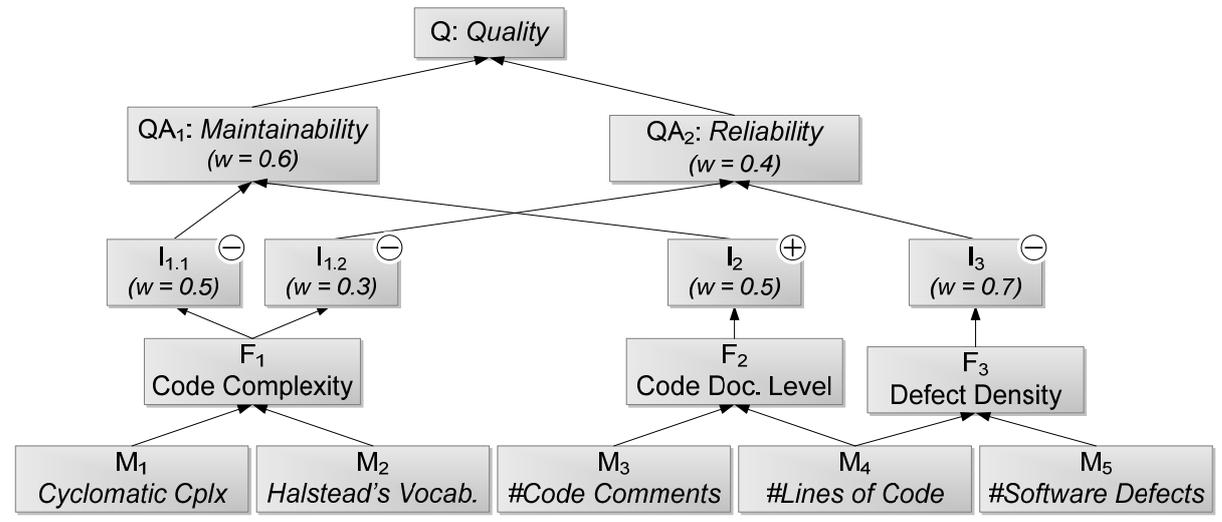

**Figure 5**: Example quality model

| Measure | M1 | M2 | M3 | M4 | M5 |
|---|---|---|---|---|---|
| Value | 5 | 50 | 500 | 1000 | 10 |

**Table 2**: Example project measurement data

***Step 2: Measure factors***. Factor $F_1$ represents an abstract property of software code – its complexity. Since both associated measures consider quite different aspects of code complexity, we decided not to combine them into an overall complexity – we left $F_1$ unmeasured. Both $M_1$ and $M_2$ will be used in step 3 for directly evaluating the impact $I_{1.1}$. The factors $F_2$ and $F_3$ represent derived measures. Their values are computed by applying simple mathematical functions: $F_2 = M_3 / M_4$ and $F_3 = M_5 / M_4$.

***Step 3: Evaluate impacts***. For each impact, we evaluate the influence of the associated factor or – if the factor's value is not available – its measures on the associated quality aspect. Since at this stage we do not really need evaluations on the 1-6 ordinal scale, we compute the values of impacts on the [0, 1] ratio scale. "Normalized" impact values can be used directly as inputs for computing the values of associated impacts in step 4.

For the impacts $I_2$ and $I_3$, we merely need to map the values of associated factors on the [0, 1] scale using an appropriate normalization function. Figure 6 illustrates two example functions for normalizing positive and negative impacts, respectively. For each value function, an acceptance threshold is defined. Up to the threshold, the factor's values map linearly onto the impact values between 0 and 1; above it, the evaluation remains constant (0 or 1 for negative or positive impact, respectively) independent of the factor's value.

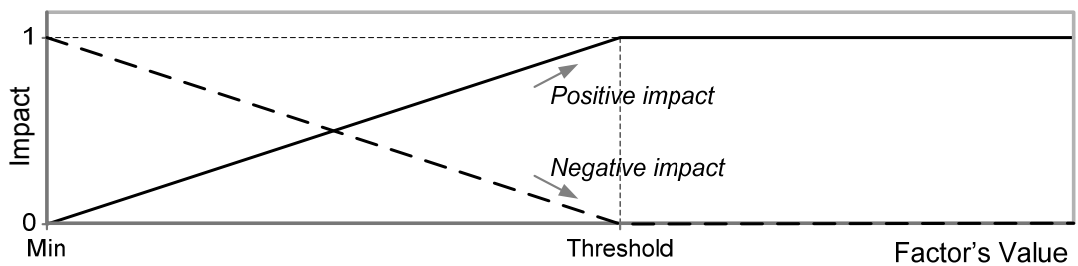

**Figure 6**: Example normalization functions

Let us assume an acceptance threshold for the code documentation level equal to 1 comment per LOC and for defect density equal to 10 defects per 1kLOC. Consequently, $I_2 = 0.5$ and $I_3 = 0.9$.

For the impacts $I_{1.1}$ and $I_{1.2}$, the value of the associated factor $F_1$ has not been computed; the values of the two measures $M_1$ and $M_2$ need to be used instead. For this purpose, we propose using the aggregation activity; that is, applying a value function (2). We assume that both measures contribute equally to both $I_{1.1}$ and $I_{1.2}$, and assign them equal weights of 0.5. Principally, we may decide on unequal weights if, for example, one of the measures has a greater impact on a particular quality aspect. Next, we need to normalize $M_1$ and $M_2$ to the range [0, 1]. For this purpose, we use the negative function illustrated in Figure 6. We set the acceptance thresholds for $M_1$ and $M_2$ to 10 and 100, respectively. This results in normalized values on $M_{1n} = 0.5$ and $M_{2n} = 0.5$. Consequently, $I_{1.1} = I_{1.2} = 0.5 \times 0.5 + 0.5 \times 0.5 = 0.5$.

***Step 4: Evaluate quality aspects***. Quality aspects are evaluated by aggregating the values of associated impacts (2) and evaluating the aggregation outcomes. The aggregated value for maintainability $V(QA_1) = 0.5 \times I_{1.1} + 0.5 \times I_2 = 0.5$; and for reliability $V(QA_2) = 0.5 \times I_{1.2} + 0.5 \times I_3 = 0.7$. For the purpose of evaluation, we define a simple evaluation function (Figure 7), which we then use for both $QA_1$ and $QA_2$. The function corresponds to the grading key used in German schools.

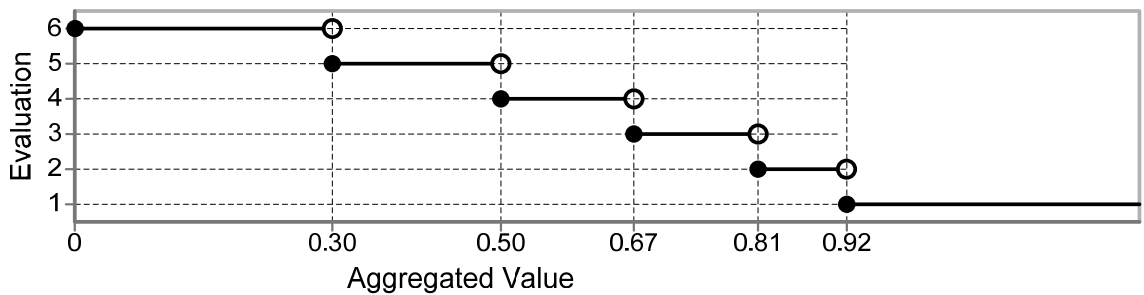

**Figure 7**: Evaluation function for quality aspects

Consequently, maintainability is evaluated at grade 4 (fair) and reliability at grade 3 (satisfactory).

***Step 5: Evaluate quality***. Finally, quality evaluation proceeds analogically to the evaluation of quality aspects. We aggregate the values of $QA_1$ and $QA_2$ using the value function (2): $Q = 0.6 \times QA_1 + 0.4 \times QA_2 = 0.58$. Next, we again use the grad-ing key defined by the evaluation function in Figure 7 to evaluate the final quality of a software product. The resulting quality is equal to 4 (fair).

| Element    | F1 | F2  | F3   | I1.1 | I1.2 | I2  | I3  | QA1 | QA2 | Q    |
|------------|----|-----|------|------|------|-----|-----|-----|-----|------|
| Value      | -  | 0.5 | 0.01 | 0.5  | 0.5  | 0.5 | 0.9 | 0.5 | 0.7 | **0.58** |
| Evaluation | -  | -   | -    | -    | -    | -   | -   | 4   | 3   | **4** |

**Table 3**: Results of example quality evaluation

## 5 Initial Validation

For the initial validation, a quality model for embedded systems software was chosen. The model was developed as part of the Quamoco project, and is conformant to the previously described meta-model. The quality aspects in this model are typical challenges faced in embedded systems, like safety and reliability. These aspects are determined by 34 factors (cf. Figure 2). The model focuses on software written in C and C++, and so PC-Lint suggests itself as measurement tool. From the rich set of available rules, 54 were chosen and classified in the quality model. The number of rule violations found in the investigated source code, normalized by lines of code, yields the measurement result. For the sake of simplicity, all elements in the quality model were assigned equal weights. The overall quality was not evaluated.

The validation was done on a set of four projects of different size (5 - 300 kLOC) written in C++ in an industrial context, which were also used to test two slightly different evaluation approaches. The goal was always an evaluation of the quality aspects based on a 6-grade ordinal scale, similar to the example evaluation above. It has to be stated that the evaluation results presented in the following are based on a very first version of a quality model that is still under development. The grades assigned are valid for use in comparisons, but do not reflect an absolute quality statement expressed in school grades.

The level of the evaluation was varied in two ways: (1) evaluation on the factor level and (2) evaluation on the quality aspect level.

The first approach was based on the idea of evaluating measurement results directly. For each measure, the number of violations was translated into a 6-grade ordinal scale based on a simple normalization function using threshold values similar to those shown in Figure 6. Depending on the thresholds, grade 6, for example, was already assigned to measures for one violation. Finally, the evaluations of the measures were aggregated into factor evaluations by simply using the average. The same was done for the aggregation of factors into quality attributes.

| Project | Average | Reliability | Availability | Resources | Maintain. | Safety | Security |
|---|---|---|---|---|---|---|---|
| Proj. 1 | 3 | 3 | 3 | 3 | 3 | 3 | 2 |
| Proj. 2 | 3 | 3 | 3 | 3 | 4 | 3 | 3 |
| Proj. 3 | 4 | 4 | 5 | 5 | 4 | 4 | 4 |
| Proj. 4 | 3 | 3 | 2 | 3 | 3 | 3 | 3 |

**Table 4**: Results of the evaluation using variant 1

A look at the results of the evaluation (Table 4) reveals that converting measurements into a 6-grade ordinal scale at the very first level of the quality evaluation and aggregating these evaluations throughout the quality model resulted in a significant loss of information. In consequence, there was only little difference in the evaluated quality of the considered software, leaving open the question of whether the used approach is able to differentiate software products with respect to their quality.

The second approach tested was similar to the example evaluation presented in section 4.3. Here, the measurement values were used to determine the degree of fulfillment of a certain factor. Each factor was given a certain, currently equal, value, to which it influences a quality aspect. Because equal weights were used for this first validation, the influence of every measure on a factor could be calculated by simply dividing the factor value by the number of assigned metrics. How much a single measure ultimately contributes to the fulfillment of a factor was determined by a normalization function similar to the one shown in Figure 6. For example, if a factor with four assigned measures has a maximum value of 1, each measure is able to contribute a maximum value of 0.25. If a measure, i.e., the number of violations of a specific PC-Lint rule is 0, then the factor receives the full value for this measure, in this case 0.25.

| Project | Average | Reliability | Availability | Resources | Maintain. | Safety | Security |
|---|---|---|---|---|---|---|---|
| Proj. 1 | 3 | 3 | 3 | 3 | 4 | 3 | 3 |
| Proj. 2 | 4 | 4 | 3 | 3 | 4 | 4 | 3 |
| Proj. 3 | 5 | 4 | 5 | 5 | 5 | 4 | 4 |
| Proj. 4 | 3 | 3 | 3 | 3 | 4 | 3 | 3 |

**Table 5**: Results of the evaluation using variant 2

For the aspect evaluation, the maximum reachable and attained factor values were summed up, put into relation, and evaluated using the function shown in Figure 7.

For example, an aspect with two impacting factors, each with a maximum value of 1 and a reached value of 0.9, has a degree of fulfillment of $1.8/2 = 0.9$ and hence, grade 2. The results of this second approach (Table 5) show greater diversity, which is a necessary prerequisite for a valid evaluation method. The analysis of the significance and appropriateness of the evaluation results, which is needed in order to improve the quality model, is part of another work package of the Quamoco project.

## 6       Summary and Future Work

This paper shows how an explicitly defined meta-model as provided by the Quamoco project can be combined with MCDA techniques to define a well-defined product quality evaluation approach. Initial application of the approach with real project data and a domain-specific quality model have shown that the diversity of the evaluation results for different products depends strongly on the way the evaluation functions are defined by the experts and on how the type of aggregation is chosen. Therefore, providing experts with clear and understandable guidelines on how to determine useful evaluation and aggregation rules is seen as a promising way for improving the usefulness of the evaluation results. Future research should focus on addressing constraints on particular MCDA and data analysis techniques. One of the directions would be to combine compensatory and non-compensatory MCDA techniques. Moreover, elements of probability theory can be considered for addressing information uncertainty issues.

### Acknowledgments

We would like to thank especially Sonnhild Namingha from Fraunhofer IESE for the initial review of the paper. This work has been partially funded by the BMBF project Quamoco (01 IS 08 023 C).